\documentclass[graybox]{svmult}

\usepackage{type1cm}        
\usepackage{makeidx}         
\usepackage{graphicx}        
\usepackage{multicol}        
\usepackage[bottom]{footmisc}

\usepackage{newtxtext}       %
\usepackage{newtxmath}       

\usepackage{color}
\usepackage{ulem}

\makeindex             

\begin{document}

\title*{Exploiting solar visible-range observations by inversion techniques: from flows in the solar subsurface to a flaring atmosphere}
\titlerunning{Inversions of solar visible-range observations}
\author{Michal \v{S}vanda, Jan Jur\v{c}\'ak, David Korda, and Jana Ka\v{s}parov\'{a}}
\institute{Michal \v{S}vanda \at Astronomical Institute, Charles University, V Hole\v{s}ovi\v{c}k\'ach 2, 180 00 Praha, Czech Republic, \at 
Astronomical Institute of the Czech Academy of Sciences, Fri\v{c}ova 298, 251 65 Ond\v{r}ejov, Czech Republic,  \email{michal@astronomie.cz} \and
Jan Jur\v{c}\'ak \at Astronomical Institute of the Czech Academy of Sciences, Fri\v{c}ova 298, 251 65 Ond\v{r}ejov, Czech Republic \and
David Korda \at Astronomical Institute, Charles University, V Hole\v{s}ovi\v{c}k\'ach 2, 180 00 Praha, Czech Republic \and 
Jana Ka\v{s}parov\'{a} \at Astronomical Institute of the Czech Academy of Sciences, Fri\v{c}ova 298, 251 65 Ond\v{r}ejov, Czech Republic}
%
%
%
\maketitle

\abstract{Observations of the Sun in the visible spectral range belong to standard measurements obtained by instruments both on the ground and in the space. Nowadays, both nearly continuous full-disc observations with medium resolution and dedicated campaigns of high spatial, spectral and/or temporal resolution constitute a holy grail for studies that can capture (both) the long- and short-term changes in the dynamics and energetics of the solar atmosphere. Observations of photospheric spectral lines allow us to estimate not only the intensity at small regions, but also various derived data products, such as the Doppler velocity and/or the components of the magnetic field vector. We show that these measurements contain not only direct information about the dynamics of solar plasmas at the surface of the Sun but also imprints of regions below and above it. Here, we discuss two examples: First, the local time-distance helioseismology as a tool for plasma dynamic diagnostics in the near subsurface and second, the determination of the solar atmosphere structure during flares. The methodology in both cases involves the technique of inverse modelling.\newline\indent}

\section{Introduction}
The Sun is our closest star and as such it serves as a prototype for the whole class of such cosmic objects \cite{2019sgsp.book.....E}. As compared to the other stars, the Sun is routinely observed from a luxurious proximity 24 hours a day, 7 days a week. The availability of the long-term near-continuous synoptic observations together with short-term high-resolution campaign observations consists a holy grail for highly detailed studies of the solar structure and dynamics. The  theories and hypotheses based on the observations of the Sun serve as constraints for theories of stellar structure and evolution. 

The Sun is an active and variable star with various phenomena observed in its atmosphere termed the \emph{solar activity}. 
These phenomena exist due to the solar interior being interwoven with the magnetic field, which takes various forms. The origin of it is still being debated in the literature  \cite{Charbonneau:2010}. Overall, the solar magnetic field depicts cyclic behaviour on a large range of time scales from seconds to millennia \cite{2017LRSP...14....3U}. There is a dynamo process running deep inside the Sun, possibly at the base of the convection zone, which recycles and strengthens the global magnetic field. Emergence of the magnetic flux tubes through the convection zone towards the surface and above forms complexes of strong magnetic field localisations, the \emph{active regions} \cite{2009LRSP....6....4F}. In active regions we typically find the most prominent phenomena of solar activity -- sunspots, prominences, and flares. All these processes are only partially understood, thus  much of the on-going and future research is dedicated to them. 

The proper understanding of the violent phenomena of solar activity, namely flares and coronal mass ejections, is necessary as these phenomena may in effect interfere with human technology, which is crucial for the quality of life as we know it \cite{2015SpWea..13..524S}. There are many open questions and without answering those we will not be able to successfully predict the solar activity, which is one of the main goals of the solar research. For instance, triggers for flares may lie in the dynamics of the plasmas at the solar surface and below and the flaring processes may strongly depend on the parameters of the atmosphere above the surface. Visible-range observations may in principle bring important information about both solar interior and atmosphere.

\section{Solar photosphere in the visible range}
In the visible-range of electromagnetic radiation, the  photosphere is the dominant source of solar radiation. The term photosphere is often replaced by ``solar surface'', but this is not very precise. Following a text-book definition, the photosphere describes the thin near-spherical shell where the visible and infrared solar continua originate \cite{2019sgsp.book...59S}. Its width is several hundreds km and the lower boundary of the photosphere is defined at the optical depth unity ($\tau_{500}=1$) -- level at which the intensity at $\lambda=500$~nm is approximately given by the continuum source function. The opacity is dominated by bound-free H$^{-}$ transitions. The formation height of the continuum radiation is wavelength dependent. For instance it is deeper in the near-infrared part of the spectrum, where free-free contributions of H$^{-}$ dominate. 

The photosphere is only several hundreds kilometers thick. One of the definitions adopts a thickness of about 300~km, which corresponds to the scale over which the gas pressure decreases by an order of magnitude. 
Even though a realistic description of the vertical stratification of the photosphere have been achieved already by 1-D models \cite{1976ApJS...30....1V}, observations are very convincing in the fact that the evaluations of the structure of the photosphere needs to be considered in all three dimensions \cite{2005ESASP.600E..12C}. Already the observations in the visible continuum by amateur telescopes reveal the pattern of granulation, the distinct convection pattern that corresponds to the thermal dissipation scale \cite{2009LRSP....6....2N}. The granular cells have a typical diameter of about 1~Mm and a life time of about 10 minutes. Both these parameters depict very wide statistical distributions. According to the current paradigm, the granules are the final products of the turbulent-convection cascade emerging from the bottom of the convection zone \cite{2019SciA....5.2307H}. Other convection-like features are present and measured in the photosphere: the mesogranules, the supergranules and possibly the giant cells. 

High-resolution observations reveal not only the granular cells, but also features of smaller size, such as magnetic bright points and other indications for the small-scale magnetic field \cite{2010ApJ...715L..26S}. In these magnetic-field concentrations the $\tau=1$ level lies even deeper, about 200~km as compared to the surrounding ``quiet'' photosphere. In these regions, the magnetic pressure contribution to hydrostatic balancing reduces the gas density. Such thin flux tubes form a magnetic network  that is preferentially located at supergranular boundaries. Even larger magnetic field concentrations create anomalous granular patterns (flux-emergence regions, ``plage'' regions) and strong enough magnetic field inhibits the convection completely by creating pores and sunspots \cite{2012LRSP....9....4S}.

Thus the transition from the 1-D vertically stratified models towards a fully 3-D time-dependent solution of (magneto)hydrodynamical equations with realistic equation of state was eminent. And it was successful \cite{2013A&A...554A.118P}. The physics behind the processes active in the quiet photosphere and closely below and above are believed to be understood quite well. That is demonstrated by the fact that the numerical simulations reproduce the observable photospheric properties remarkably well \cite{2009LRSP....6....2N}. Detailed simulations reproduce even specialties such as elongated granules in the regions with emerging magnetic fields or particular flows around spots and pores \cite{2009Sci...325..171R,2010ApJ...720..233C}.

\section{Sub-surface dynamics}
From the analyses of solar observations together with proper modelling based on the theory of stellar structure and evolution it turns out that the Sun is a main-sequence star with an effective temperature of $5778\pm3$~K, radius of $(6.960\pm0.001)\times 10^8$~m and a mass of $(1.9889\pm0.0003)\times10^{30}$~kg  \cite{2004suin.book.....S}. Because of the action of gravity, such a star has a significant radial stratification, generally showing three main layers. The central region up to some 25\% of radius is occupied by the \emph{core}, where the physical conditions allow the thermonuclear fusion to run mostly by means of the proton-proton chain. This is the region where the solar luminosity, equivalent of almost 4$\times$10$^{26}$~W, originates. The central temperature reaches 15.7~MK and the pressure of 2.5$\times$10$^{16}$~Pa according to the state-of-the-art models  \cite{1996Sci...272.1286C}. 

At around 25\% of solar radius the temperature of the ambient plasma decreases below the value of around 7~MK. The efficiency of the thermonuclear reactions is very low henceforth, however the state parameters of the plasma are such that the plasma is fully ionised and the opacity is very low for the high-energy photons originating from the proton-proton chain. Due to the temperature gradient the energy is transported towards the surface via radiation, photons diffuse by free-free radiative scattering. This layer is called \emph{the radiative zone}. 

At 71\% of the solar radius  \cite{1997MNRAS.287..189B,2004ApJ...614..464B} the mean plasma temperature decreases under $\sim$2~MK and recombination of ions starts to occur locally. The opacity rapidly increases due to the bound-free radiative absorption of photons by heavy elements. The energy transport via radiation is no longer possible. The convection sets in and begins to be the main energy transport agent until the surface, through the whole \emph{convection zone}. The ever-changing granulation is a consequence of the convective overshoot from the convection zone underlying the photosphere. The convection zone is a principal layer, where the dynamo is seated somewhere and from where the flux tubes later forming sunspots and other phenomena rise up. 

Studying  the convection zone is not straightforward. The sub-surface of the Sun is optically thick, preventing us from directly observing the interior layers. Understanding the properties of the plasma in these regions has consequences for  theories of convection, stability of sunspots, the dynamics of stratified convection, and others. Most of current knowledge about convection comes primarily from computational work, e.g.  \cite{2005AA...429..335V,2006ASPC..354...92B,2009ApJ...691..640R}. Helioseismic inversions of the sub-surface flows play an important role in constraining these theories.  

The physics of convection in the low-viscosity, large-density and large-temperature gradient regime associated with the convection zone is not well known  \cite{2014SoPh..289.3403H}. Most of the convection zone is moderately vertically stratified, but as the plasma approaches the photosphere, it undergoes a rapid expansion because of the steep near-surface density gradient. With the rapid decrease of temperature in the near-surface layers ionisation zones of several elements (e.g. helium and hydrogen) form, and the release of the latent heat by the recombination is thought to power the various scales of turbulent convection. 

\subsection{Oscillations}
A powerful way of imaging the solar interior is via inferences gathered from studying the statistics of the acoustic and surface gravity waves detectable at the surface. Solar pressure and surface gravity modes are generated randomly by the vigorous turbulence in the upper convection zone. These oscillations are best observed in the solar photosphere by measuring Doppler shifts of photospheric absorption lines. 

\begin{figure}
    \includegraphics[width=0.5\textwidth]{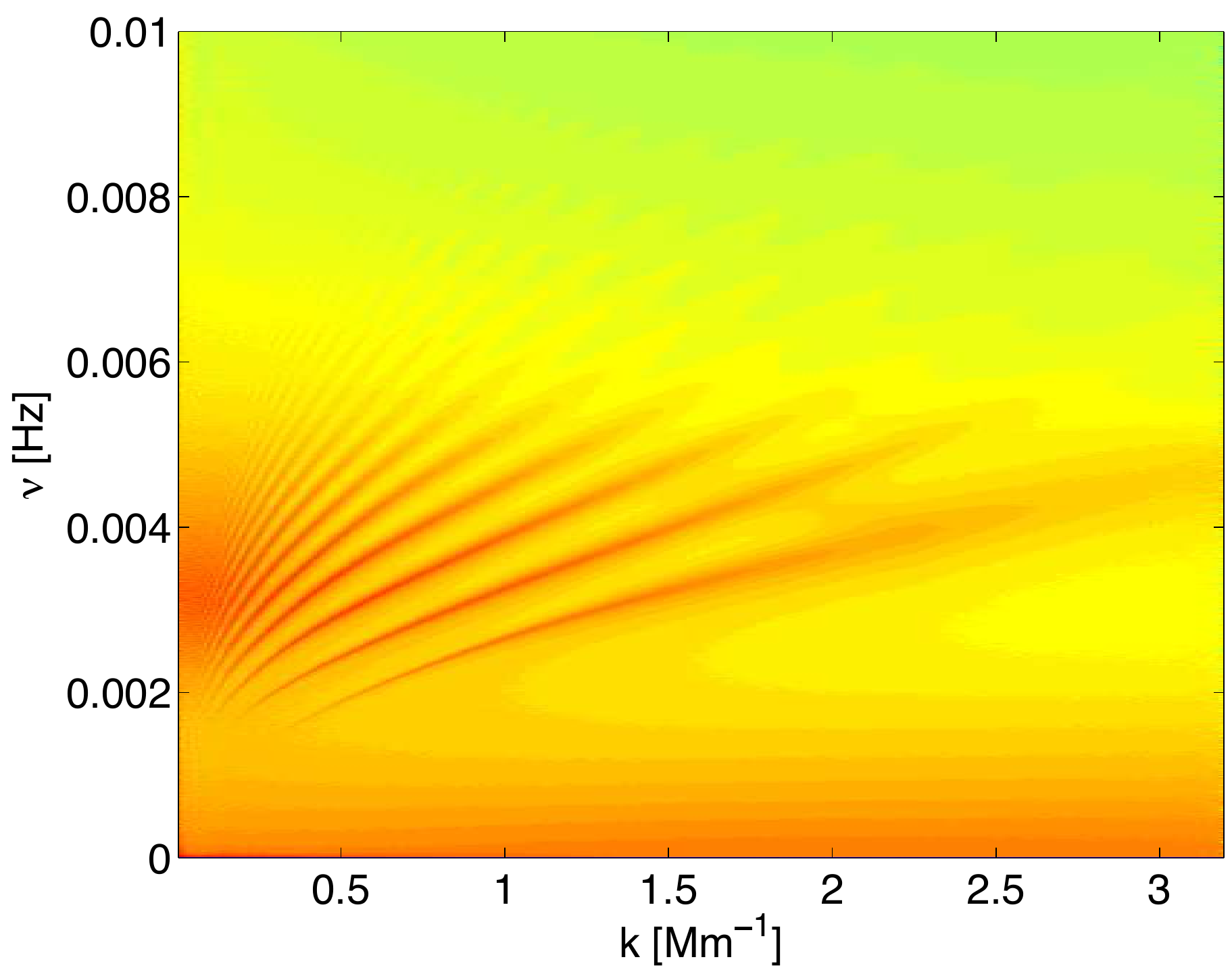}
    \caption{An example $k-\nu$ diagram as constructed from 24-hour Dopplergram series recorded by Helioseismic and Magnetic Imager (HMI) on-board of Solar Dynamics Observatory (SDO). Signal of low frequency corresponds to the convection zone. The lowest positioned ridge is the $f$-mode ridge, others are $p$-mode ridges. ``Five-minute oscilations'' with frequency around 3.3~mHz are result of interference of $p$-modes. A lack of signal above 5.3~mHz is related to an acoustic cut-off frequency. The colour scale in arbitrary units indicates the spectral power with greenish colours having a low power and reddish colours having a large power.}
    \label{pic:k-w}
\end{figure}

The spatio-temporal power spectrum of solar oscillations provides the best evidence for the existence of the resonant standing waves. These resonant modes form the structure of non-intersecting ridges in a $k-\omega$ diagram, where $k$ stands for the wave number and $\omega = 2\pi\nu$ for the cyclic frequency of the waves. The position of the ridges is a very sensitive indicator of the internal structure of a star. An example of the $k-\nu$ diagram is in Fig. \ref{pic:k-w}. One can see obvious signatures of resonant modes depicted by ridges of increased spectral power.

A vast majority of the modes seen in the solar $k-\omega$ diagram belong to the group of $p$-modes, which are acoustic modes where the pressure is a restoring force. The $p$-modes propagate mostly in a convectively unstable layers, that is in the near-surface convection zone. The ridge positioned lowest on the $k-\omega$ diagram is a $f$-mode ridge. $f$-modes are surface gravity modes, similar to the oscillations of the free surface. Frequencies and wave numbers (which relate to the wavelengths) are examples of helioseismic observables. 


\subsection{Helioseismology}
Changes in the internal structure of the Sun, including localised changes, induce shifts in frequencies of the modes. Forward modelling allows us to relate anomalies like flows, thermal hot/cold spots etc. to changes in helioseismic observables. It is not straightforward to infer the characteristics of these anomalies from observables. Sophisticated techniques of inverse modelling -- or ``inversions'' in short -- must be adopted.

The aim of helioseismic inversions is to reveal the structure of the subsurface flows (rotation, meridional circulation, convection), magnetic fields, and to measure deviations in the plasma state parameters (temperature, density, pressure) from a quiet Sun average. 

Localised perturbations in the parameters of solar plasmas may be investigated using the approach of local helioseismology. The time--distance helioseismology \cite{1993Natur.362..430D}, one of the local helioseismic methods, has proven to be very useful. In recent years, time--distance helioseismology has been used to determine near-surface flows  \cite{2000JApA...21..339G,2000SoPh..192..177D,2004ApJ...603..776Z,2008SoPh..251..381J}, flows beneath sunspots  \cite{1996Natur.379..235D,2001ApJ...557..384Z,2006ApJ...640..516C,2008SoPh..251..291C,2009SSRv..144..249G,2009arXiv0912.4982M} and flows in their vicinity  \cite{2000JApA...21..339G}, and to study the rotational gradient at the base of the convection zone  \cite{2009ApJ...693.1678H}, etc. 

\begin{figure}
    \centering
	\includegraphics[width=0.5\hsize]{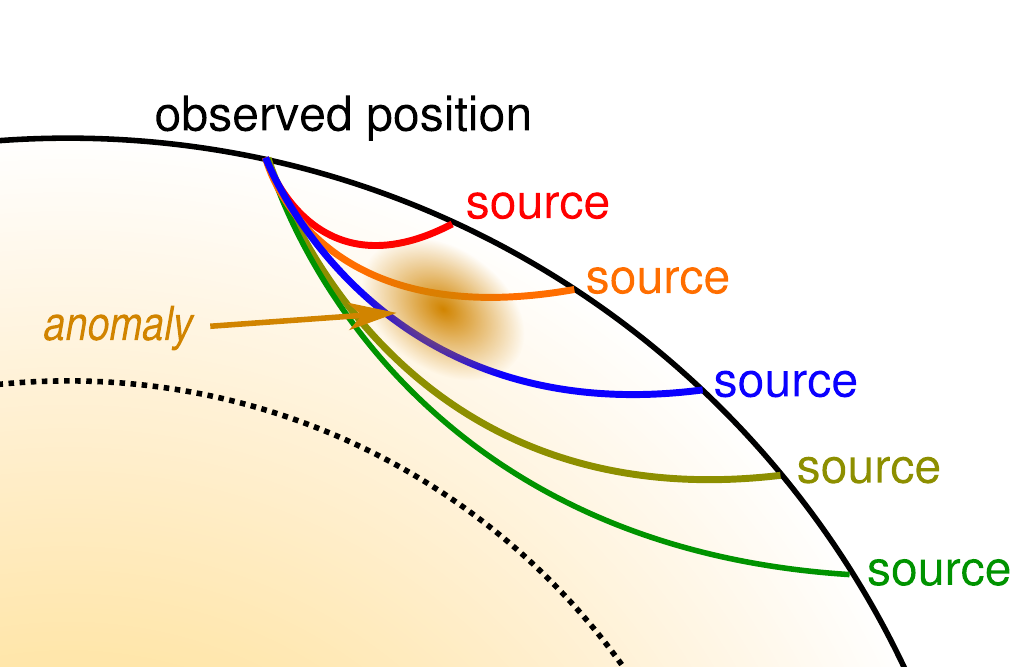}
	\caption{Schematic representation of the time--distance technique. The seismic waves are excited at various sources and travel through the solar interior to the observed position. Different colour indicate different modes (with different wavelengths). Some of these ray travel through an anomaly in the interior, which affects the total travel time of the affected waves. From combining the travel-time measurement of various modes one can learn about the nature of the anomaly.}
	\label{pic:TD}
\end{figure}

Time--distance inversions (see a review  \cite{2010arXiv1001.0930G} and a scheme in Fig.~\ref{pic:TD}) are based on a concept of a perturbed travel time $\delta\tau$,
\begin{equation}
    \delta\tau = \tau_{\rm obs}-\tau_{\rm model}.
\end{equation}

The observed travel time $\tau_{\rm obs}$ is measured from observations (usually from a series of intensity or Doppler-shift measurements of the photospheric lines) by means of a time-domain cross-correlation of the signal between two different points on the solar surface. The model travel time $\tau_{\rm model}$ is computed from the assumed background solar model. The background model is usually assumed to be vertically stratified with no magnetic fields and no plasma velocities, such as Model S  \cite{1996Sci...272.1286C}. 

The actual state of the solar interior at a certain point may depart from the background model, i.e., state quantities may be perturbed or flows may be present. The usual assumption used in a vast majority of helioseismic studies is that these perturbations of the plasma state are small compared to the background model. Thus we may assume a linear perturbation and solve the linearised equation of the wave propagation.  

The perturbed travel time $\delta\tau$ can theoretically be computed from the model as

\begin{equation}
\delta \tau \left(\boldsymbol{r}\right)= \int \limits_{\odot} \mathrm{d}^3 \boldsymbol{x} \boldsymbol{K}(\boldsymbol{x}) \cdot \delta \boldsymbol{q}(\boldsymbol{x}). 
\label{eq:dtau}
\end{equation}
where $\boldsymbol{r}$ indicates the horizontal position vector and $z$ the vertical coordinate in a plane-parallel approximation of the 3-D position vector $\boldsymbol{x}=(\boldsymbol{r},z)$. Here $\boldsymbol{K}$ is a vector of \emph{sensitivity kernels} which represents the linear sensitivity of the travel-time measurements to the perturbations $\delta\boldsymbol{q}$  \cite{2004ApJ...614..472G,2007AN....328..228B}. The sensitivity kernels are defined by the background model and the level of approximation (ray approximation, Born approximation, etc.). The vector of perturbations may contain quantities such as flows, speed of sound, density, etc. Since the solar oscillations are excited by vigorous surface convection, in reality the measured travel time contains a large random-noise realisation component, which needs to be taken into account in further analyses. 

The goal of an inverse problem is to invert Eq. (\ref{eq:dtau}) to determine $\delta\boldsymbol{q}$ from the measurements of $\delta\tau$. Such an inversion is usually not possible due to the presence of random noise. 

In the Sun, a full spectrum of waves with different properties is excited. Individual modes propagate more-or-less independently throughout the solar interior, thus by measuring the travel times of different modes we obtain independent information about the nature of the solar interior. By filtering, it is possible to separate distinct modes of the waves and measure their travel times. Consequently, for each filtered wave mode there is a separate equation (\ref{eq:dtau}), therefore also the travel times may symbolically be written in the vector form as $\delta \boldsymbol{\tau}$. On the other hand, all these equations share the same vector of perturbers $\delta \boldsymbol{q}$. In helioseismic inversions one combines all independently measured travel times to learn about the perturbers. 

Helioseismic inversions are usually performed by using two principal methods: The regularised least squares (RLS) and optimally localised averaging (OLA). The RLS method  \cite{1996ApJ...461L..55K} seeks to find the models of the solar interior, which provide the best least-squares fit to the measured travel-time maps, while regularising the solution (e.g., by requiring the smooth solution). The OLA method was developed for geoseismology  \cite{1968GeoJ...16..169B,1970RSPTA.266..123B}. 
Subtractive-OLA (SOLA) method \cite{1992AA...262L..33P}, a form suitable for use in helioseismology, is based on explicitly constructed spatially confined averaging kernels by taking linear combination of sensitivity kernels, while simultaneously keeping the error magnification small. The resulting coefficients are then used to linearly combine the travel-time maps and obtain an estimate for structure and magnitude of solar plasma perturbations. A SOLA-type inversion is the principal method demonstrated here. The SOLA has been used in time--distance local helioseismology in the past  \cite{2007AN....328..234J,2008SoPh..251..381J}, where the ability of SOLA inversions to reveal the structure of 3-D internal flows was demonstrated. An efficient approach to solve fully consistent SOLA inversions was has been also introduced \cite{fastOLA}.

\subsection{SOLA time--distance inversions}

\subsubsection{The basic principle}
Unlike the RLS approach, the SOLA methodology does not minimise the difference between the measured travel times and those modelled from the modified solar model. SOLA assumes that the inverted estimates of the quantities may be obtained using a linear combination of the travel-time maps, symbolically written for $\alpha$-th component $\delta q_\alpha$ of the vector $\delta\boldsymbol{q}$ as
\begin{equation}
\delta q_{\alpha}^{\mathrm{inv}} (\boldsymbol{r}_0) = \boldsymbol{w}(\boldsymbol{r}-\boldsymbol{r_0}) \cdot \delta\boldsymbol{\tau} (\boldsymbol{r}),
\label{eq:deltaq}
\end{equation}
where $\boldsymbol{w}$ are the inversion weight functions.  From an oversimplified point of view, SOLA performs the deconvolution of the travel-time maps. 

Taking into account a noise term (the measured travel times are noisy) and using Eqs.~(\ref{eq:dtau}) and \ref{eq:deltaq}), we obtain
\begin{equation}
\delta q_{\alpha}^{\mathrm{inv}} ( \boldsymbol{r}_0) = \int \limits_{\odot} \mathrm{d}^3 \boldsymbol{x}' \boldsymbol{\mathcal{K}} (\boldsymbol{r}' - \boldsymbol{r}_0,z') \cdot \delta \boldsymbol{q} (\boldsymbol{x}')+\mathrm{noise}\ ,
\label{eq:inv_by_akern}
\end{equation}
where the object $\boldsymbol{\mathcal{K}}$ is referred to as the (vector) \emph{averaging kernel}. It is defined as 
\begin{equation}
    \boldsymbol{\mathcal{K}} \left(\boldsymbol{r}', z; z_0\right) = \int  \mathrm{d}^2 \boldsymbol{r}\, \boldsymbol{w} \left(\boldsymbol{r} - \boldsymbol{r}_0; z_0\right) \cdot \boldsymbol{K}\left(\boldsymbol{r}' - \boldsymbol{r}, z\right)
    \label{eq:akern}
\end{equation}
and quantifies the level of spatial smearing of the real quantity $\delta\boldsymbol{q}$. 

In the SOLA method we search for the weights $\boldsymbol{w}$ so that the vector averaging kernel $\boldsymbol{\mathcal{K}}$ resembles the target function $\boldsymbol{\mathcal{T}}$. The target function is an user-selected initial estimate for the averaging kernel which indicates the desired localisation in the Sun. An example of commonly used target function is a 3-D Gaussian with a peak at a chosen target depth $z_0$. 

In practise the SOLA cost function $\chi^2_{{\rm SOLA}}$ is minimised, which can be written in a simplified form
%
\begin{equation}
\chi^2_{{\rm SOLA}} = \left|\left|\boldsymbol{\mathcal{K}} (\boldsymbol{x}')- \boldsymbol{\mathcal{T}} (\boldsymbol{x}') \right|\right|^2 + \mu \sigma_\alpha^2 + \nu \sum \limits_{\beta \neq \alpha} \left|\left|\mathcal{K}_{\beta} (\boldsymbol{r}')\right|\right|^2+\chi^2_{\rm other}. 
\label{eq:chi}
\end{equation}

The first term $\left|\left|\boldsymbol{\mathcal{K}} (\boldsymbol{x}')- \boldsymbol{\mathcal{T}} (\boldsymbol{x}') \right|\right|^2$ of the right-hand side of Eq. (\ref{eq:chi}) then evaluates the misfit between the averaging kernel and a target function $\boldsymbol{\mathcal{T}}$. 

The second term of Eq. (\ref{eq:chi}) is the regularisation of the random-noise level in the inverted quantity,
\begin{equation}
\sigma^2_{\alpha} = \boldsymbol{w}^{\rm T} \boldsymbol{\Lambda} \boldsymbol{w},
\end{equation}
where $\boldsymbol{\Lambda}$ is the travel-time noise covariance matrix. 
The matrix $\boldsymbol{\Lambda}$ may be obtained either from the reference model or from the travel-time measurements. 

\subsubsection{The cross-talk}
The third term $\sum_{\beta \neq \alpha} \left|\left|\mathcal{K}_{\beta} (\boldsymbol{r}')\right|\right|^2$ in Eq. (\ref{eq:chi}) is related to the cross-talk and regularises the non-diagonal components (that is all components except for the inverted $\alpha$-th) of the averaging kernel. The meaning of the cross-talk is illustrated by Eq. (\ref{eq:inv_by_akern}). There, the inverted estimate of the $\alpha$-th component is composed not only from the appropriate $\delta q_\alpha$ smoothed with the respective averaging kernel ${\mathcal K}_\alpha$, but also from the ``irrelevant'' perturbers $\delta q_\beta$ for $\beta \ne \alpha$ smoothed with their respective averaging kernels ${\mathcal K}_\beta$. 

The physical meaning of the averaging kernel is demonstrated in Figs. \ref{pic:vx_0.5} and \ref{pic:rakern_x_4.0}. According to Eq. (\ref{eq:inv_by_akern}) the inverted quantity ($v_x^{\rm{inv}}$ in this case) is obtained by cross-correlation of the averaging-kernel vector and vector $\delta \boldsymbol{q}$ of the real physical quantities. For each position $\boldsymbol{x}_0 = \left(\boldsymbol{r}_0, z_0\right)$, the weighted average of real quantity $\delta q_{\beta}$ is computed where the weights equal to $\cal{K}_{\beta}$. The final $v_x^{\rm{inv}}$ is then a sum of the noise term and individual weighted averages for all $\beta$s (columns one to four in Fig. \ref{pic:vx_0.5}). In the case of $v_x$ inversion in Fig. \ref{pic:rakern_x_4.0} one can see that $v_x^{\rm{inv}}$ is mostly given by real $v_x$ averaged over around 15~Mm in horizontal and 4~Mm in vertical direction (see contours in Fig. \ref{pic:rakern_x_4.0}). The other contributions are small because $\mathcal{K}_y$, $\mathcal{K}_s$, $v_z$, and $\delta c_s$ are small, thus products of $\mathcal{K}_{\beta}\, \delta q_{\beta}$ are small for these off-diagonal components (cf. individual contributions in Fig. \ref{pic:vx_0.5}). 

To complete the discussion of the terms in the inversion cost function we add that $\chi^2_{\rm other}$ contains additional terms such as the regularisation of the weights or the constraint on the normalisation of the averaging kernel. User-given trade-off parameters $\mu$ and $\nu$ balance the terms in the cost function. 

The framework seems to be well established, but some unknowns remain that need to be cleared out. For instance, the inverse problem is cast in a set of algebraic equations with a system matrix that is ill-posed and usually nearly singular. The numerical errors from calling the mathematical pseudo-inverse of the system matrix may propagate throughout the rest of the computations. The consistency of the numerical implementation must be evaluated carefully. 



\subsubsection{Interpretation of the inversion}
The description above gives an indication that there are three different quantities that are needed to be determined in order to interpret the time--distance inversion results properly. 

\begin{itemize}
    \item \emph{Inverted map} $\delta q_\alpha^{\rm inv}$, which gives the inverted estimates for the physical quantity $\delta q_\alpha$ at positions $\boldsymbol{r}_0$ and at target depth $z_0$.
    
    \item \emph{Noise level} $\sigma_\alpha$ gives the root-mean-square value of the random-noise component in the inverted map. The particular noise-component realisation in the inversion is not known and is an inseparable part of the inverted map. 
    
    \item Vector \emph{averaging kernel} $\boldsymbol{\mathcal{K}}$, defined by Eq. (\ref{eq:akern}), gives an important information about the smearing of the inverted quantity. It also gives the real depth sensitivity (whereas the value of $z_0$ is the target depth which only serves as a parameter). In Fig. \ref{pic:akern_h_vz} one can see the depth dependence of averaging kernels of four inversions with different target depths $z_0=(0.5,2,4,6)~{\rm Mm}$. A careful reader can see that the peak sensitivity of these kernels is different from their indicated target depths. 
    
    The averaging kernel also contains information about the presence of the sidelobes. Sidelobes are places with non-negligible values of the averaging kernel which are at larger distances from the central point. Such an averaging kernel is not well localised (see an example in Fig. \ref{pic:rakern_z_4.0}). Thus, the inverted estimate in a given point is spoiled by the properties of plasma which is far away and which possibly has different properties.
    
    The off-diagonal components for $\alpha \ne \beta$ indicate the leakage of unwanted perturbations to the inverted map, the cross-talk.
    Ideally these cross-talk components of the averaging kernel should be negligible, so that the inverted map is not polluted by the cross-talk.
\end{itemize}

\subsubsection{Lessons learnt from the validation}
 One way to check the consistency of the inversion modelling is a validation using known inputs. These known inputs may come from the realistic simulation of solar convection. Such a validation then gives limits that could be expected from the inversions of the real-Sun observables \cite{2019A&A...622A.163K}. 



\begin{figure}
	\centering
	\includegraphics[width=\textwidth]{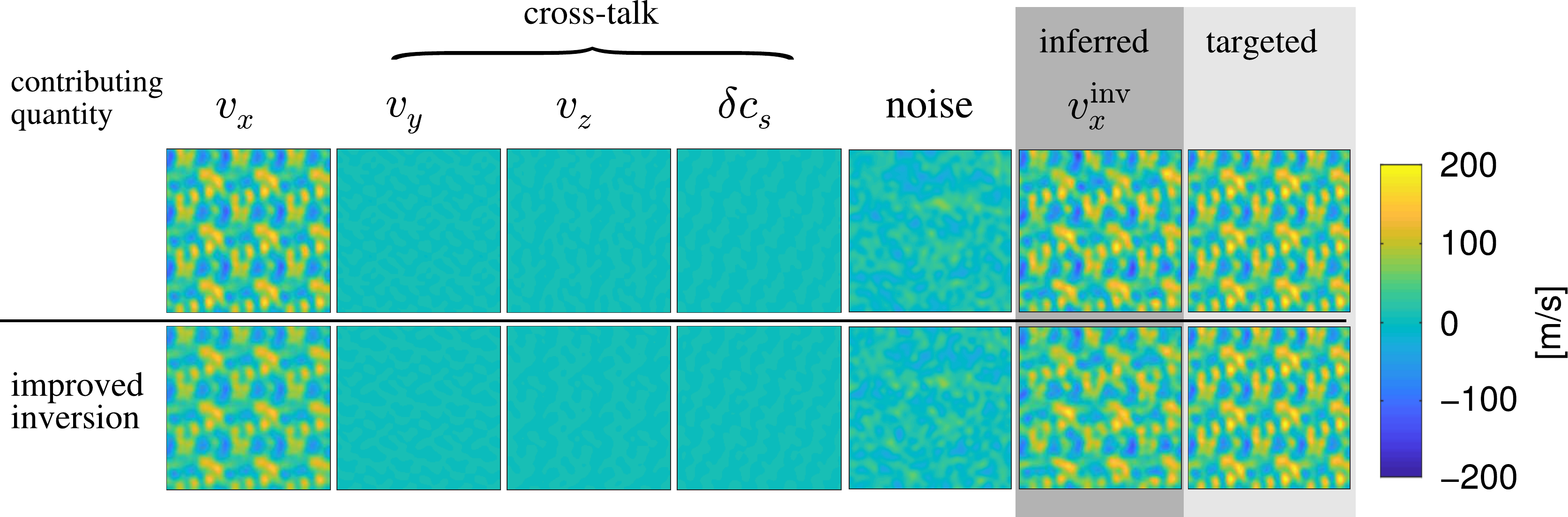}
	\caption{Top row: inversion without minimisation of the cross-talk. Bottom row: inversion with minimisation of the cross-talk. From left, columns one to five show individual contribution to the inverted $v_x^{{\rm inv}}$. The sixth column is $v_x^{{\rm inv}}$ and in the seventh column there is an ideal answer. The extent of both the horizontal and vertical axes is approximately 290 Mm. Compare with Fig.~\ref{pic:vz_crosstalk}}.
	\label{pic:vx_0.5}
\end{figure}

The inversions for the horizontal flow components ($v_x$ and $v_y$) are well-posed in the near-surface layers. The results of the inversion are in a great agreement with the known flows which come from a simulation of the convection zone. In Fig. \ref{pic:vx_0.5} one can see individual contributions to the inverted $v_x^{{\rm inv}}$. From the left, first five columns correspond to individual terms in Eq. (\ref{eq:inv_by_akern}), the sixth column is a sum of individual contributions (left-hand side of Eq. (\ref{eq:inv_by_akern})) and the right-most box shows an  ideal answer of our inversion (noiseless smearing of the real quantity by the inversion target function). In the first row, minimisation of the cross-talk was not applied, while in the second row, it was applied. The differences in corresponding columns are negligible. Thus, horizontal flow is a strong perturber and possible contributions from the remaining weak perturbers (the vertical flow and the sound-speed perturbations) are negligible even when the cross-talk is not minimised.

\begin{figure}
    \includegraphics[width=\textwidth]{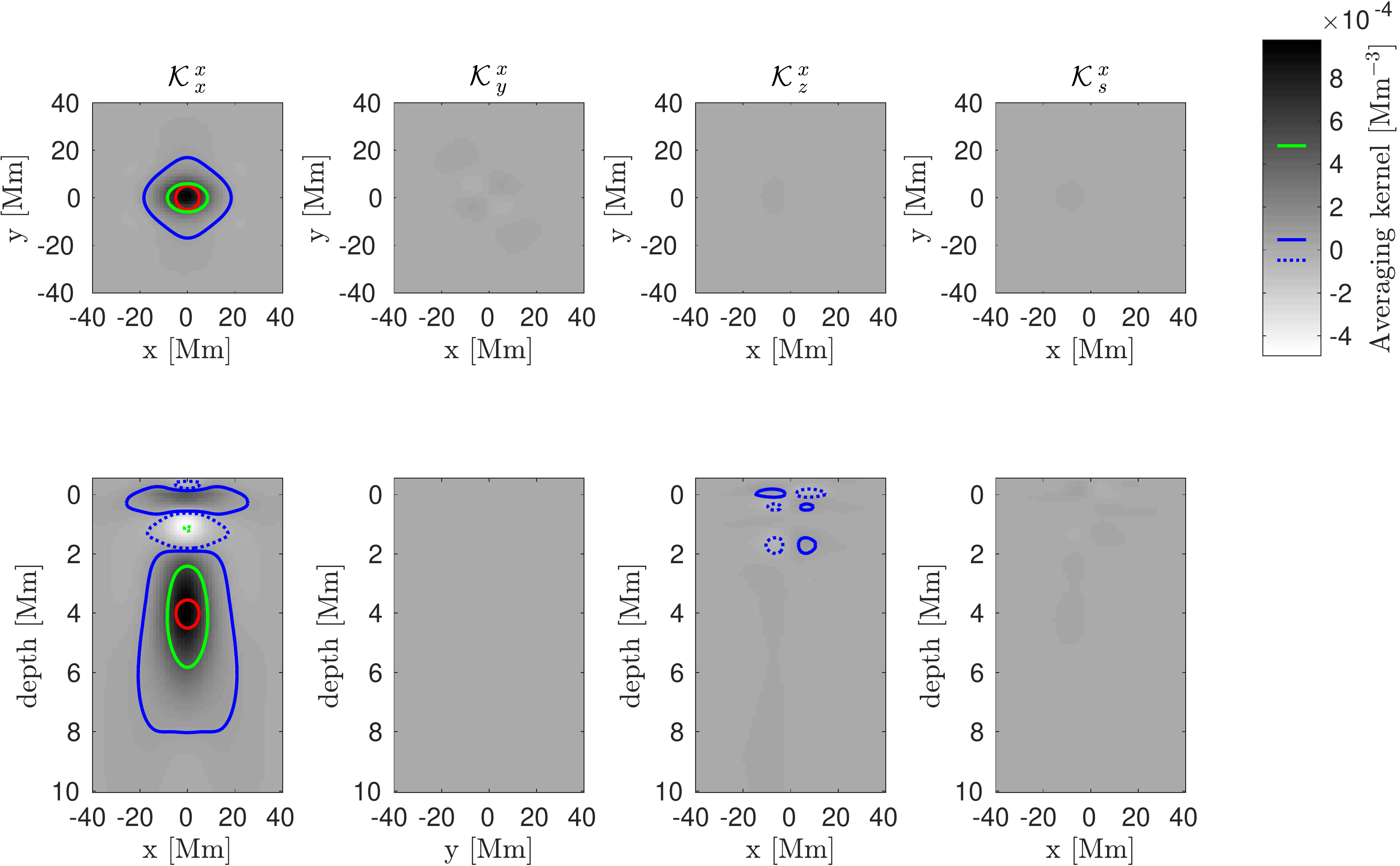}
    \caption{The averaging kernel for $v_x$ inversion at 4.0~Mm depth. The columns indicate the contributions from the individual quantities considered in the inversion, the terms not in the direction of the inversion indicate the leakage of the other quantities -- the cross-talk. The red curve corresponds to the half-maximum of the target function at the target depth. The solid green and dotted green curves correspond to plus and minus of the half-maximum of the averaging kernel at the target depth, the blue solid and blue dotted lines correspond to $+5\%$ and $-5\%$ of the maximum of the averaging kernel at the target depth, respectively. In the top row, there are horizontal slices of the averaging kernel at the target depth and in the bottom row, there are vertical slices perpendicular to the symmetries.}
    \label{pic:rakern_x_4.0}
\end{figure}

With the travel times averaged over 24 hours or so, it is possible to invert for the horizontal flow snapshot on supergranular scales roughly in the upper 5~Mm of the convection zone. An example of the averaging kernel of such an inversion is shown in Fig. \ref{pic:rakern_x_4.0}. For deeper inversions different approaches must be taken. One should either increase the level of averaging (by increasing the extent of the averaging kernel) or invert for a statistical representation. In the latter case the ensemble averaging over a set of representatives of the same kind (e.g. set of individual supergranules) is a suitable approach  \cite{2012ApJ...759L..29S,2014SoPh..289.3421D}.

\begin{figure}
\centering
	\includegraphics[width=0.50\textwidth]{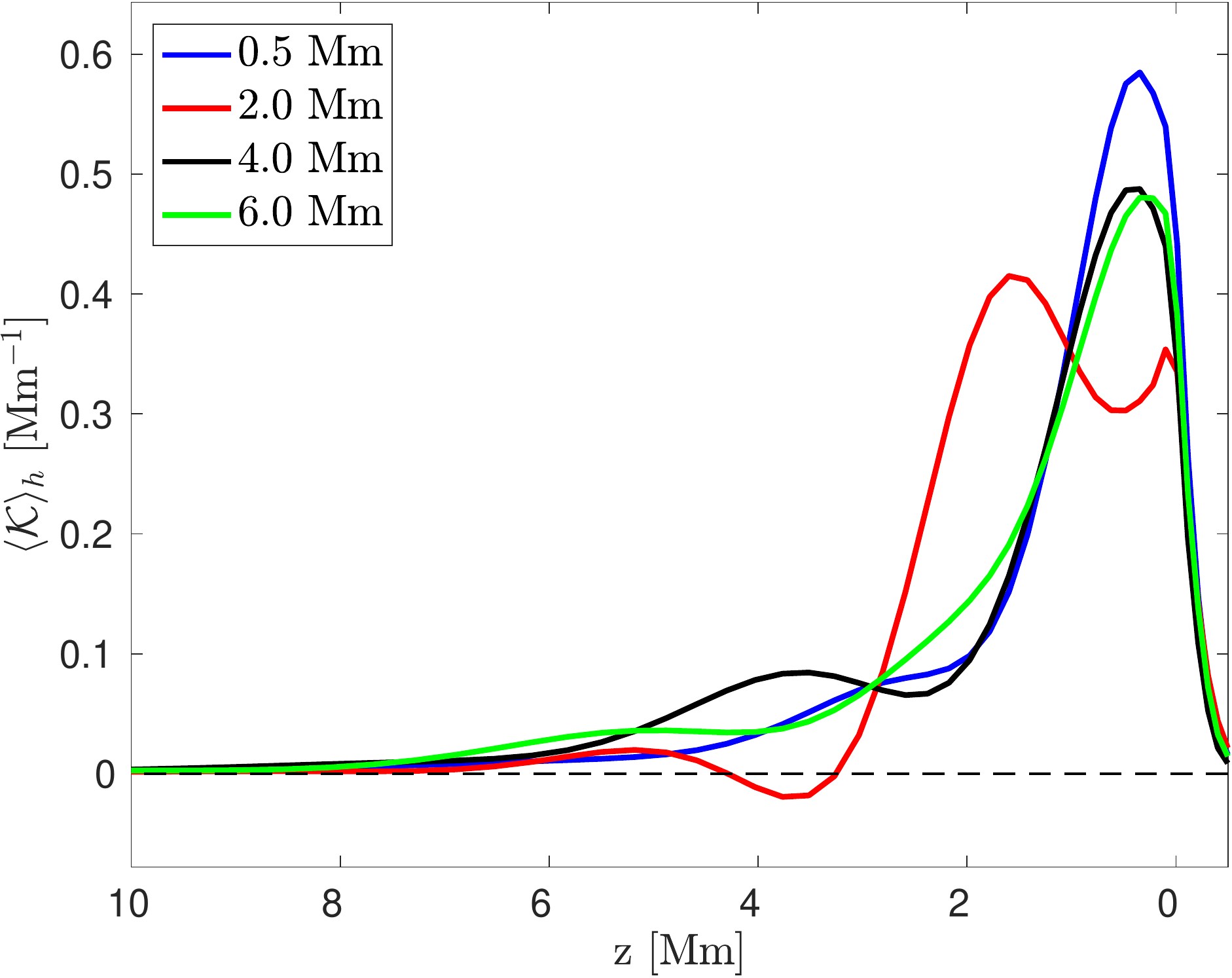}
	\caption{Examples of the horizontal integrals of the averaging kernels  for $v_z$ inversion as a function of depth. }
	\label{pic:akern_h_vz}
\end{figure}

\begin{figure}
    \includegraphics[width=\textwidth]{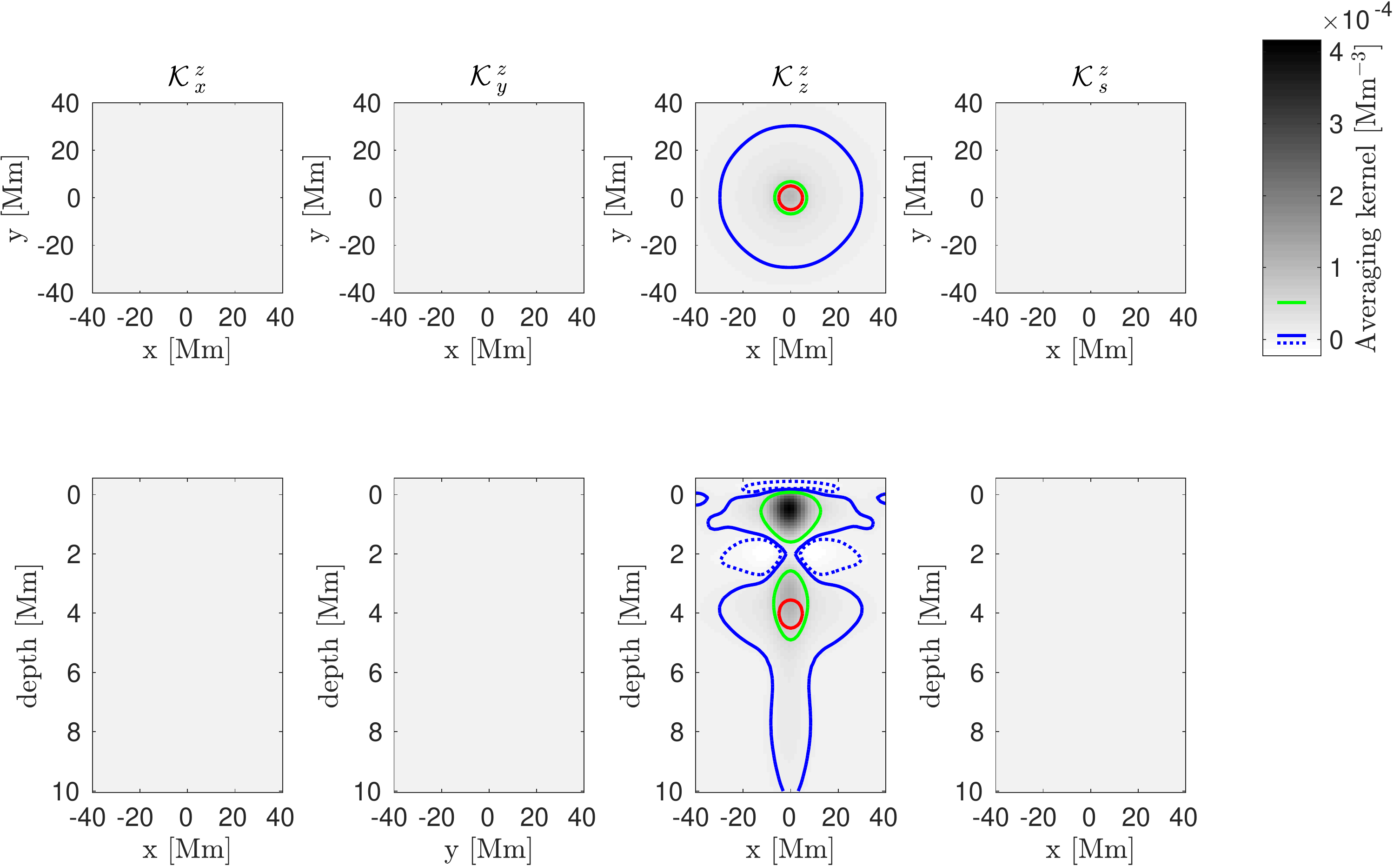}
    \caption{The averaging kernel for $v_z$ inversion at 4.0~Mm depth; however, the sensitivity of the inversion is mostly around 1~Mm below the surface. See Fig. ~\ref{pic:vx_0.5} for  the definition of contours.}
    \label{pic:rakern_z_4.0}
\end{figure}

\begin{figure}
\centering
	\includegraphics[width=0.95\textwidth]{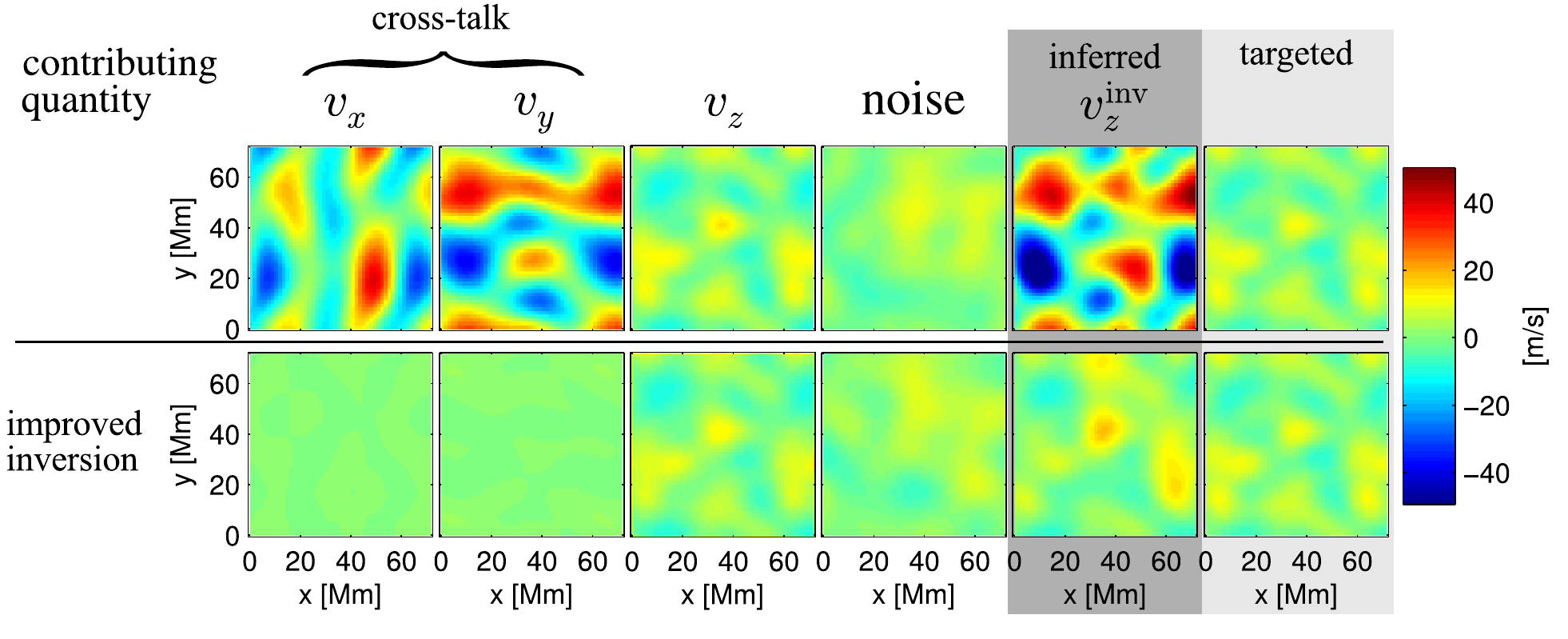}
	\caption{All components of an example $v_z$ inversion at 1~Mm depth. Top row with the cross-talk ignored, bottom row with the cross-talk minimised. We demonstrate that if cross-talk is not addressed, horizontal components will leak into the inverted $v_z$ and cause a bias. Obviously, inferred estimate is then in the anticorrelation with the expectations. }
	\label{pic:vz_crosstalk}
\end{figure}

For the successful inversion for the vertical flow component ($v_z$) it is essential to minimise the cross-talk, otherwise the results are 
significantly distorted  
by the leakage mainly from the horizontal components of the flow; the inverted vertical flow is almost in anticorrelation with the expected output  \cite{2007ApJ...659..848Z,Svanda_2011}, see Fig.~\ref{pic:vz_crosstalk}. The leakage from the sound-speed perturbations is acceptably small. For the travel times averaged over 24 hours or so it is not possible to invert for the flow snapshot deeper than around 1~Mm. In Fig. \ref{pic:rakern_z_4.0} there is the averaging kernel for $v_z$ inversion at 4.0~Mm depth. It is evident that the averaging kernel is localised around 1~Mm below the surface, not around targeted 4~Mm (compare to equivalent averaging kernel for the horizontal flow component in Fig.~\ref{pic:rakern_x_4.0}). The depth sensitivity of several inversions is nicely seen in Fig. \ref{pic:akern_h_vz}, where horizontal integrals of averaging kernels for $v_z$ inversions in different depths are plotted. Even though the target depth is changing from 0.5 Mm to 6.0 Mm, the depth sensitivity of the inversion is nearly the same. Attempts to perform deeper inversions lead either to a noise level much larger that the signal, or to a strong departure from the target function (as in this case)  \cite{Korda_Svanda_Zhao_2019}. Attempt to invert for the snapshot of the vertical flow on supergranular scales with a signal-to-noise ratio larger than one naturally converges towards the shallow ``surface'' inversion. 

\begin{figure}
	\includegraphics[width=\hsize]{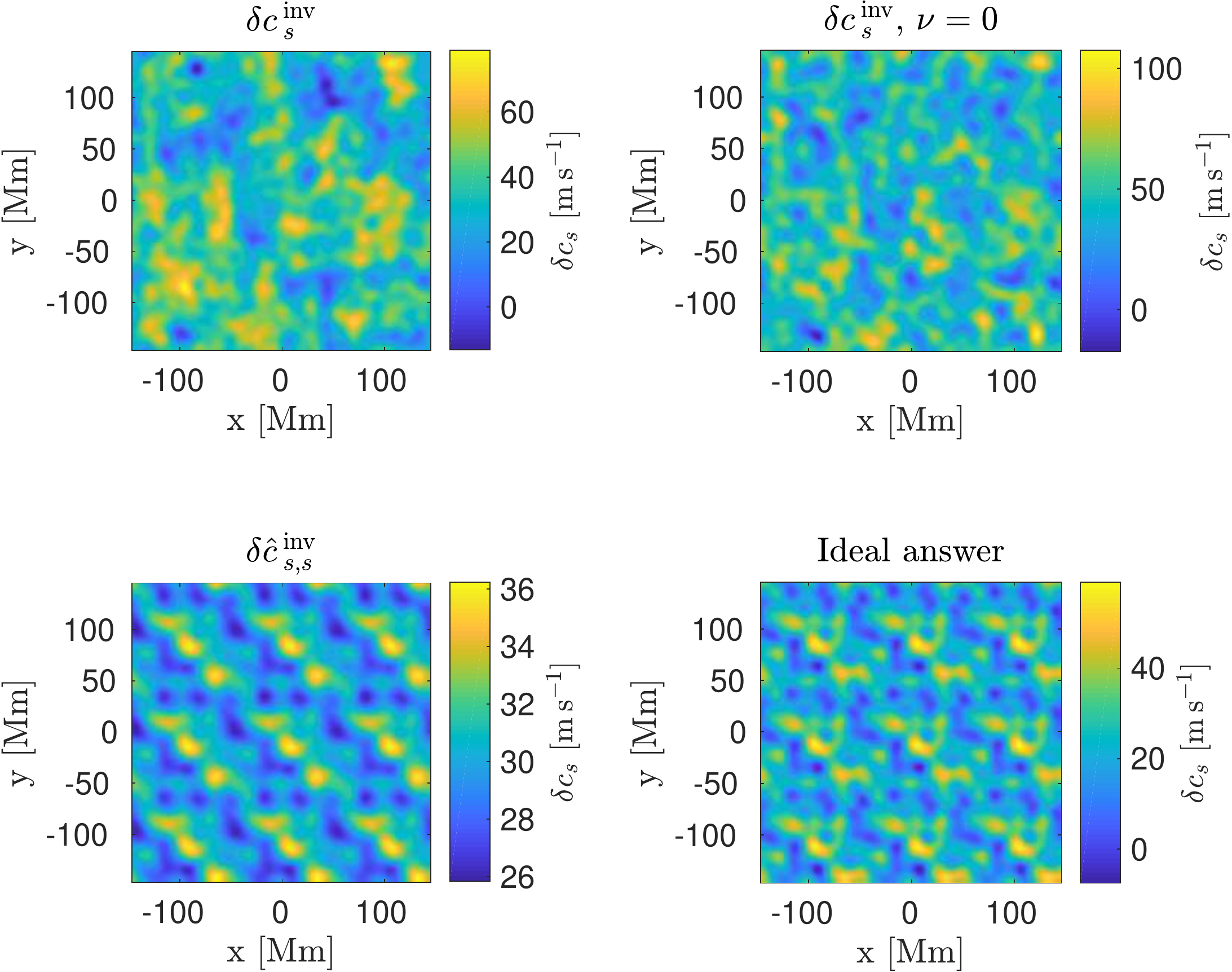}
	\caption{Top row: $\delta c_s^{\mathrm{inv}}$ with minimisation of the cross-talk and $\delta c_s^{\mathrm{inv}}$ without minimisation of the cross-talk. Bottom row: contribution in the direction of inversion and the ideal answer.}
	\label{pic:cs_0.5}
\end{figure}

The realistic perturbation of the speed of sound is also a weak perturber. Thus the cross-talk minimisation is essential. Unlike in the case of the vertical flow, in the case of the sound-speed perturbations the cross-talk is positively correlated. Thus the cross-talk ``helps'' to ``measure'' $\delta c_s$. Such an approach leads to an overestimation of the magnitude of the sound-speed variations by almost a factor of two in the near-surface layers. Such an inversion, however, is not the inversion for $\delta c_s$. An example of inverted sound-speed perturbation in near-surface layers is plotted in Fig. \ref{pic:cs_0.5}. The first row shows the inverted maps with and without cross-talk minimisation (similar to the fourth column in Fig. \ref{pic:vx_0.5}). In the second row there are, from left to right, individual contribution to inverted result with the cross-talk minimisation (similar to the fourth column in the bottom row in Fig. \ref{pic:vx_0.5}) and the ideal answer.

\vspace{14pt}
\noindent By using the local helioseismology, one can learn about the dynamics of the solar subsurface. The knowledge of this dynamics is important for assessment of the evolution of the magnetic field. Unfortunately, helioseismic inversions do not seem to help in constraining the subsurface structure of the magnetic field in the active regions due to a very strong interaction of various kinds of waves in the magnetised plasma  \cite{2009arXiv0912.4982M}. 

\section{Information about the higher atmosphere}
The coupling between the solar sub-surface and the structures of the magnetised atmosphere is striking  \cite{2009SSRv..144..317W}. For instance, a particular flow topology may contribute in triggering the solar flares. In a sudden eruption of a quiet-Sun filament followed by a solar flare found that the zonal shear flow possibly contributed to triggering the filament eruption and the consequent flare ignition \cite{2008A&A...480..255R,2018A&A...618A..43R}. 

The upper atmospheric layers have a very low density and it is difficult to speak about them in term of the spherical shells. The chromosphere is a warped and highly dynamic surface that is thin around kilogauss concentrations of the magnetic field and thick in regions where acoustic shocks appear in otherwise cool internetwork gas. Higher up the chromospheric fibrils are structured by the magnetic field that extend from the network and from plage regions. The chromosphere is very dynamic, where we find evidence for magnetic reconnections, Alfv\'{e}n waves, magnetically guided converted acoustic waves and many more which form the chromospheric structures. The chromosphere loses much of the energy by radiation in strong resonance lines of hydrogen, helium, calcium and magnesium. 

These higher layers do not contribute much in the visible range to the total solar irradiance, because the photosphere overshines them. However, locally there are phenomena, which may outshine even the background photosphere.

\subsection{White-light flares}

Solar flares are widely believed to be a consequence of reconnection of the coronal magnetic field in a peculiar configuration. The magnetic energy stored in the entangled coronal loops is released suddenly during the flare and a large portion of the flare energy is radiated away in a wide range of wavelengths emerging from the intensively heated flare atmosphere. 

In the visible range of wavelengths the usual line emission is often accompanied by enhancement of continuum radiation and such flares are called \emph{white-light flares} (WLF,  \cite{1966SSRv....5..388S,1989SoPh..121..261N}). There are various mechanisms proposed for enhancement of the optical continuum: hydrogen bound-free and free-free transitions, Thomson scattering, and H$^{-}$ emission. Futhermore, each mechanism may dominate in different atmospheric layers spanning from the photosphere through the temperature minimum region to the chromosphere, and all require an increase in temperature and electron density in those layers.

\begin{figure}
    \centering
	\includegraphics[width=0.65\hsize]{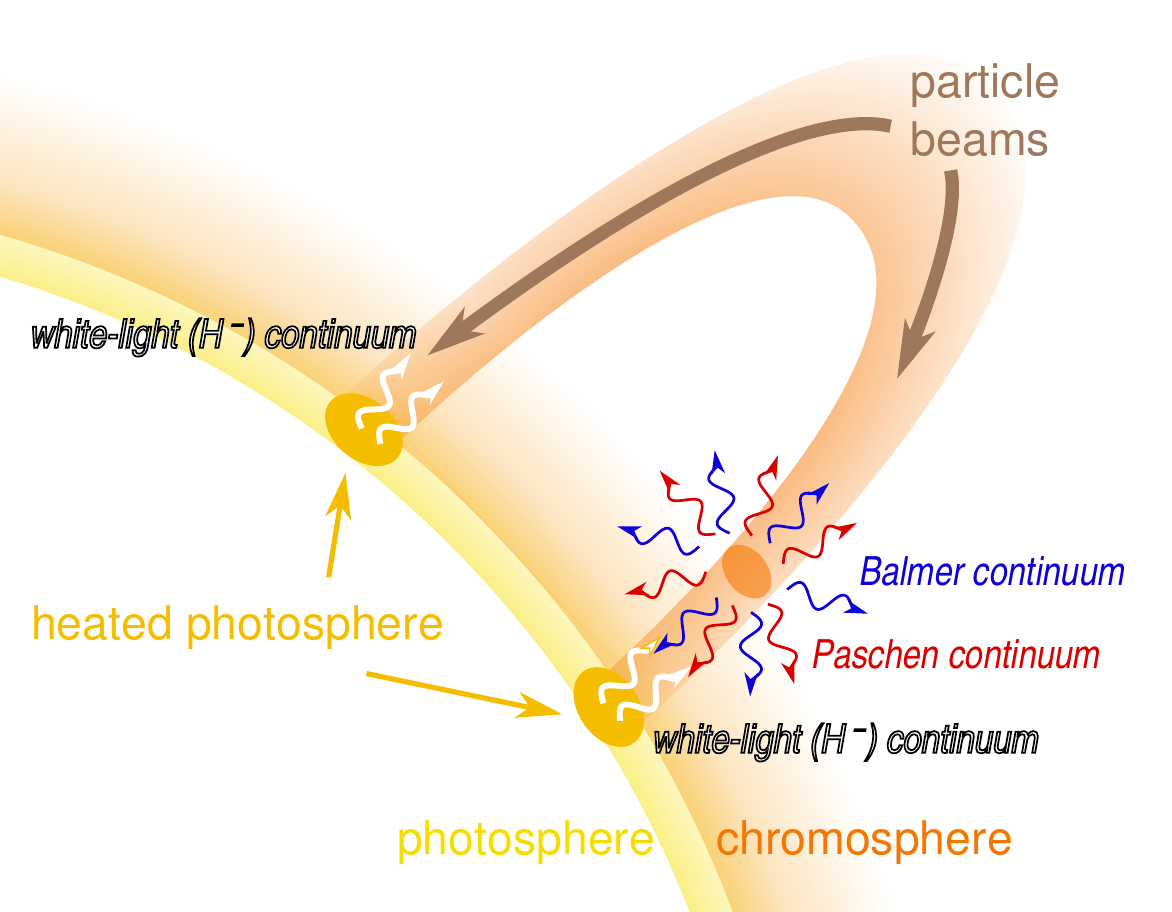}
	\caption{Schematic model of two possible origins of the white-light flare. In the left leg of the coronal loop the model of heating of the photosphere by particle beams is considered. In the right leg the schematic representation of the backwarming model is considered. }
	\label{pic:WLF}
\end{figure}

However, it is still debated how these layers are heated (see Fig.~\ref{pic:WLF}). Several processes have been proposed:  electron and/or proton bombardment, XEUV heating, Alfv\'{e}n wave dissipation, etc.  \cite{1990ApJ...365..391M}. Moreover, it has been shown that the photosphere and the chromosphere can be radiatively coupled via photospheric heating by  H$^{-}$ absorption of the hydrogen Balmer continuum, which originates in the chromosphere. This backwarming then can lead to increased photospheric (H$^{-}$) radiation  \cite{1989SoPh..124..303M}.

To disentangle the contributions to the visible-light continuum, specific observations  and dedicated models are needed. Combining flare observations of several photospheric and chromospheric lines together with visible-light continuum and non-LTE modelling, i.e. allowing departures from local thermodynamical equillibrium (LTE), a semi-empirical model of a WLF was constructed \cite{1990ApJ...360..715M}. Recently, using off-limb flare observations in the SDO/HMI pseudo-continuum and non-LTE RHD approach  \cite{Heinzel:2017}, the presence of the hydrogen Paschen continuum originating in the chromosphere was reported. Other HMI off-limb sources related to a flare were found to be of two kinds (chromospheric and coronal), and interpreted as free-bound continuum (and possible line emission) and Thomson scattering, respectively  \cite{Martinez:2014,Saint-Hilaire:2014}. Furthermore, the hydrogen Balmer continuum was observed during flares by IRIS  \cite{2014ApJ...794L..23H, 2016ApJ...816...88K,2017ApJ...836...12K}. Additionally, near-infrared emission at $1.56 \mu$m was detected during several X-class WLFs  \cite{2006ApJ...641.1210X,2012ApJ...750L...7X}. In  an undisturbed solar atmosphere this emission is considered to originate at the opacity minimum located  below $\tau_{500}=1$.

\subsection{Spectral lines and their inversion}

It is the goal of solar spectroscopy to determine reliably various physical parameters in the solar atmosphere. Spectral profiles of polarized light, i.e., the four Stokes profiles $I$, $Q$, $U$, and $V$, contain a lot of information about the conditions in the line-forming region. In the solar photosphere, the shape of polarized light profiles of magnetically sensitive lines is typically governed by Zeeman effect. Thus observations of such lines allow us to determine also the properties of the magnetic field, a physical parameter that plays an important role in many processes of solar activity.

To obtain estimates of the physical parameters in the solar photosphere, one of the fundamental steps is to solve the radiative transfer equation (RTE) for polarized light 
 \begin{equation}
    \frac{{\rm d}\mathbf{I}}{{\rm d}\tau} = \mathbf{K} (\mathbf{I}-\mathbf{S}),
    \label{RTE}
 \end{equation}
where $\mathbf{I}=[I, Q, U, V]^T$ is the Stokes vector of polarized light, $\mathbf{S}$ is the source function, and $\mathbf{K}$ is the so-called propagation matrix. In the solar photosphere, which is sufficiently dense and collisional rates are high enough, we can assume LTE. This implies that the source function is not a source of polarized light and is used in a form $\mathbf{S}=[B_\nu(T), 0, 0, 0]^T$, where $B_\nu$ represents the Planck function. Propagation matrix  $\mathbf{K}$ contains absorption, dichroism, and dispersion coefficients that are dependent on the physical parameters of the atmosphere. 

The formal solution of Eq. (\ref{RTE}) can be written in the form 
 \begin{equation}
    \mathbf{I}(0)=\int_0^\infty {\rm d}\tau\, \mathbf{O}(0,\tau) \mathbf{K}(\tau) \mathbf{S}(\tau),
    \label{RTE-fs}
 \end{equation}
where $\mathbf{I}(0)$ is the outgoing intensity of polarized light, $\tau$ is the optical depth, and $\mathbf{O}$ is the evolution operator. It can be shown that the evolution operator must fulfill the following equation \cite{1985SoPh...97..239L}: 
 \begin{equation}
    \frac{{\rm d}\mathbf{O}(\tau,\tau')}{{\rm d}\tau}=\mathbf{K}(\tau)\mathbf{O}(\tau,\tau').
    \label{evolution operator}
 \end{equation}
 
Eq. (\ref{evolution operator}) is the reason why there is no general analytical solution of RTE for polarized light. Solution in the form of attenuation exponential is valid only in specific cases because the matrices do not commute in general. The specific case is called Milne-Eddington (ME) atmosphere and it fulfills two necessary conditions. First, the propagation matrix $\mathbf{K}$ is independent on optical depth, i.e., the physical parameters defining the matrix elements are constant with optical depth. Second, the source function depends linearly on optical depth.  As the temperature determines the shape of the source function via the Planck function, the assumption of the linear dependence of source function on the optical depth determines the temperature stratification, which is not realistic. If more realistic and thus more complicated  models of atmosphere are used, the RTE is solved numerically. 

For any given model atmosphere the RTE (\ref{RTE}) is solved and the resulting intensities of polarized light are hereafter denoted as $\mathbf{I}^\mathrm{syn}$. The goal is to estimate a model atmosphere that produces $\mathbf{I}^\mathrm{syn}$ as similar as possible to the observed intensities of polarized light $\mathbf{I}^\mathrm{obs}$, i.e., to minimize the cost function
 \begin{equation}
    \chi^2=\frac{1}{\nu} \sum\limits_{k=1}^4 \sum\limits_{i=1}^M \left[ I_k^{\rm obs}(\lambda_i)-I_k^{\rm syn}(\lambda_i) \right]^2,
    \label{chi2}
 \end{equation}
where the internal sum is performed over $M$ wavelength samples indicated by index $i$, the external sum is performed over individual Stokes profiles, and the value of $\nu$ corresponds to the number of degrees of freedom, that is the difference between the number of observables and the number of free parameters of the model.
 
There are two principal methods to minimise $\chi^2$. First, compute $\mathbf{I}^\mathrm{syn}$ from a huge grid of atmospheric models and search for the best match with the $\mathbf{I}^\mathrm{obs}$ within the grid. Examples how to do it quickly using the method of the principal component analysis were published \cite{2000A&A...355..759R, 2001ApJ...553..949S}. Second, use a model atmosphere and alter the physical parameters of the model until the $\chi^2$ is minimized.  

The second method is widely used for spectral lines formed in the solar photosphere, because of the LTE conditions  be computed in a reasonable time even if RTE (\ref{RTE}) is solved numerically. Apart from the LTE assumption, the inversion codes also assume that the model atmosphere is in a hydrostatic equilibrium. Although this assumption is only rarely fulfilled in solar atmosphere, it does not influence directly the resulting values of free parameters of the inversion (magnetic field strength, inclination and azimuth, line-of-sight velocity). The assumption of hydrostatic equilibrium allows us to compute the pressure from the temperature stratification directly. The pressure is further used to derive parameters like density and electron pressure that is a necessary parameter for the line synthesis in the SIR code (Stokes Inversion based on Response functions). These parameters influence the opacity of the atmosphere and thus the height scale of the obtained model atmosphere, but the Stokes profiles of photospheric lines are typically not sensitive to these physical parameters. 

For huge datasets and automatic data reduction, the inversion methods usually assume Milne-Eddington atmosphere, the synthesis of Stokes profiles is computed analytically, and therefore these codes are very fast. Spectropolarimetric observations from satellites are automatically inverted using ME codes, e.g., VFISV code designed for SDO/HMI data  \cite{2011SoPh..273..267B} or MERLIN code\footnote{https://www2.hao.ucar.edu/csac/csac-data/sp-data-description} used to Hinode/SOT data. The disadvantage of ME codes is a very simplified model atmosphere that cannot produce asymmetric Stokes profiles. However, with increasing spatial resolution of the telescopes, the asymmetries of the Stokes profiles are commonly observed and more complicated atmospheric models have to be assumed to explain them.
 
A very powerful method of spectral-line inversion was introduced in  \cite{Cobo:1992}. This inversion code SIR (Stokes Inversion based on Response functions) allows for realistic temperature stratification and also the magnetic field vector and the macroscopic line-of-sight velocity can change with height in the atmosphere. The methodology is very flexible so it allows to invert many spectral-line profiles at once, thereby improving the resulting inverted estimates of the physical parameters. 

\begin{figure}
    \centering
    \includegraphics[width=\textwidth]{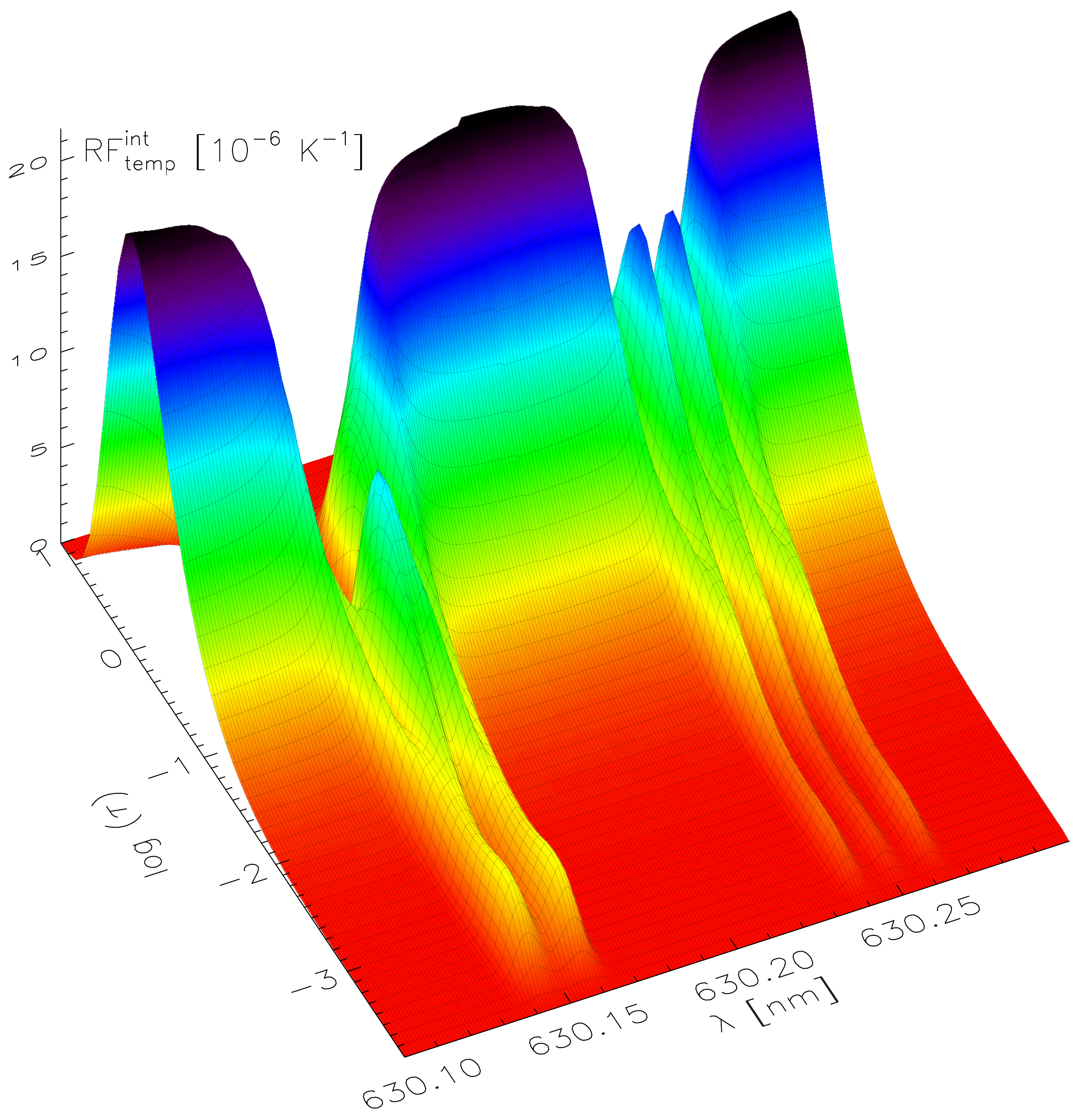}
    \caption{Response function of the Stokes $I$ profile to the temperature perturbations for the pair of neutral Fe lines around 630~nm. It illustrates that the continuum intensity is determined by temperature around $\log \tau = 0$ and the line wings and cores are more sensitive to temperature changes at higher atmospheric layers.}
    \label{RF_temp}
\end{figure}

The method is based on the concept of \emph{response functions}, which come from the linearisation of the RTE (\ref{RTE}). The response function $\boldsymbol{R}(\lambda,\tau)$, where $\lambda$ is the wavelength and $\tau$ is the optical depth, describes a linear relation between the atmospheric parameters and the emergent intensity. 

\begin{equation}
    \delta \boldsymbol{I}(\lambda)=\int\limits_0^\infty \boldsymbol{R}(\lambda,\tau) \delta x(\tau) \mathrm{d}\tau, 
    \label{eq:SIRdeltaI}
\end{equation}
where $\delta \boldsymbol{I}$ represents the modification of the emergent Stokes spectrum $\boldsymbol{I}$ which is caused by the perturbation $\delta x$ in a single physical parameter for simplicity. Vector $\boldsymbol{R}$ is a vector of response functions corresponding to the Stokes-vector components and tells us how the observed spectrum responds to the modifications of the physical conditions in the model. Response functions behave the same way as partial derivatives of the spectrum with respect  to the physical quantities.  Within  linear approximation, response functions give the sensitivities  of the emerging Stokes profiles to perturbations  of plasma parameters in the given model atmosphere. Example of response function of the Stokes $I$ profile to the temperature perturbation in a model atmosphere corresponding to the sunspot umbra is shown in Fig.~\ref{RF_temp}. Here one can see the sensitivity of the pair of iron lines to temperature at different atmospheric levels.

In an iterative process the atmosphere is modified in such a way that the cost function (\ref{chi2}) is minimized. In a linear approximation the cost function of the altered atmospheric model $\chi^2 (\vec{x} + \delta\vec{x})$ can be approximated by the Taylor series of the cost function of the original atmospheric model $\chi^2 (\vec{x})$
\begin{equation}
    \chi^2(\vec{x} + \delta\vec{x}) \simeq \chi^2(\vec{x}) + \delta\vec{x}^{T} (\nabla\chi^2 + \mathbf{H}'\delta\vec{x}),
    \label{chi2_tailor}
\end{equation}
where $\nabla\chi^2$ are partial derivatives of the cost function and can be derived from the response functions, $\mathbf{H}'$ is so-called curvature matrix containing the second partial derivatives of $\chi^2$, i.e., $H'_{ij}=1/2 \; \partial^2 \chi^2 / \partial x_i \partial x_j$. The elements of the curvature matrix are non-trivial functionals of the response functions. 

The inversion code SIR \cite{Cobo:1992} uses the Marquardt algorithm to minimize the cost function. Following Eq. (\ref{chi2_tailor}), the goal is to solve the equation  
\begin{equation}
    \nabla\chi^2 + \mathbf{H}'\delta\vec{x} = 0,
\label{marquardt1}
\end{equation}
i.e., the model atmosphere condensed in the vector $\vec{x}$ is altered in a way that its modification $\delta\vec{x}$ fulfills the equation
\begin{equation}
    \delta\vec{x} = -\mathbf{H}'^{-1} \nabla\chi^2.
\label{marquardt2}
\end{equation}
Note that the change of the atmospheric model is in the code further controlled by an additional parameter, with its value based on the evolution of the cost function during the iterative process of the spectral line fitting. In each iteration the response functions need to be calculated again to ensure that the final solution in the last step fits well to the linear approximation from the previous step. The problem of radiative transfer of polarised light and the spectral line inversions is described in detail in \cite{Iniesta:2003}.

The inversion code SIR is widely used by the community but there are other inversion codes that allow for stratification of plasma parameters with height in the atmosphere and are also based on linearisation of the RTE (\ref{RTE}), see e.g.  \cite{2000A&A...358.1109F, 2012A&A...548A...5V}.

\subsection{Atmosphere in the flare}
The SIR code \cite{Cobo:1992} is designed to invert spectral lines for  situations where non-LTE effects are not important, thus its use is usually limited to the investigation of the structure of the magnetised photosphere. 
However, there are case studies, including the example below \cite{2018A&A...620A.183J}, where some information about the upper atmospheric levels may also be inferred. 
In the following we describe a special application of SIR code to spectropolarimetric measurements of photospheric lines obtained during a solar flare.

The Solar Optical Telescope (SOT,  \cite{Tsuneta:2008}) aboard the Hinode satellite  \cite{Kosugi:2007} has provided observations in broad-band filters as well as spectropolarimetry of a pair of photospheric Fe I lines. Its continuum broad-band filters have been used to detect and analyse WLF emission  \cite{2014ApJ...783...98K}. Here, we focus on spectropolarimetric data. The spectropolarimeter attached to the SOT has been measuring the Stokes profiles of the Fe I 630.15 and 630.25~nm lines since 2006. Analyses of these data have given us great insight into the structure of the magnetic field in the solar photosphere, both in quiet-Sun and active regions. In all cases, these Fe I lines were observed in absorption on the solar disk. This is caused by a temperature decrease with height, and thus a decrease in the source function in the layers of the solar photosphere where these lines are formed. Only in the case of observations at the extreme solar limb, were these lines  detected in emission  \cite{Lites:2010}.

\begin{figure*}[!t]
 \centering \includegraphics[width=0.85\linewidth]{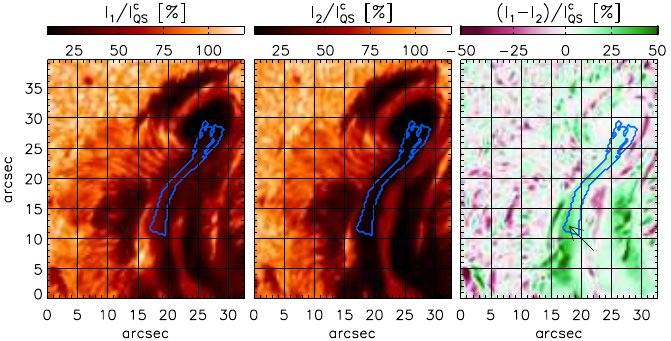}
 \caption{Continuum intensity maps reconstructed from the two Hinode raster scans. The left map was scanned between 11:57~UT and 12:04 UT and captured a WLF ribbon; the middle map was scanned between 12:19~UT and 12:42~UT. On the right is the intensity difference between these two scans. The blue contour indicates the region where we ascribe the intensity difference to the WLF. The arrow points to a pixel where we observed the Stokes profiles displayed in Figs.~\ref{profile_comparison} and~\ref{profile_reduction}. The labels  $I_1$ and $I_2$  correspond to the continuum intensities observed during the first and second Hinode/SP scan, respectively.}
 \label{hinode_int}
\end{figure*}

In September 2017 the raster scan of  Hinode spectropolarimeter (SP) crossed a WLF ribbon and captured a unique
set of emission profiles that allow us to study in detail the response of the solar photosphere to an X9.3-class flare (SOL2017-09-06T11:53, see also \cite{2018SpWea..16.1261G}). 
Continuum intensity maps of Hinode rasters are shown in Fig.~\ref{hinode_int} and
the blue contour indicates a region where the intensity enhancement is caused by the WLF ribbon and not by the evolution of the sunspot fine structure.

The above mentioned code SIR was used to determine the physical properties of the solar photosphere. To account for the complex emission profiles observed in the flare ribbon, the temperature was allowed to change at several (five) optical depths $\tau$. 

\begin{figure*}[!t]
 \centering \includegraphics[width=0.85\linewidth]{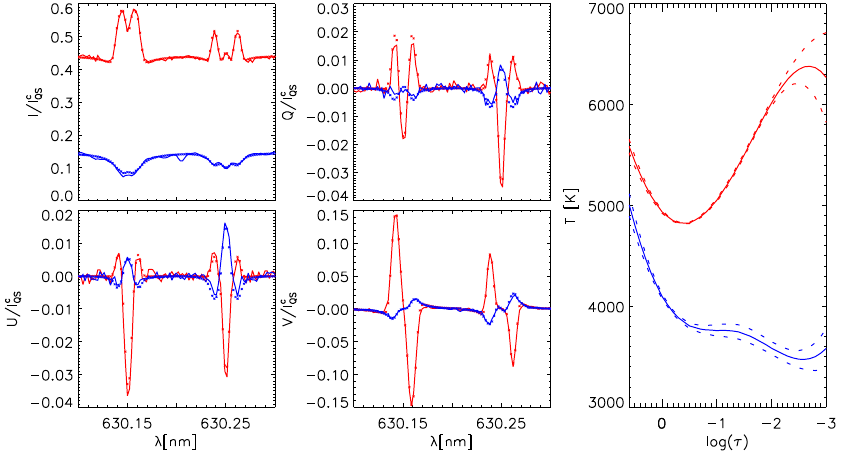}
 \caption{Left, small plots: Comparison of the Stokes profiles observed during the WLF (red lines) and in the post-WLF phase (blue lines); the $*$ symbols in the respective colours indicate the best fit of these profiles achieved with the inversion code. Right: Temperature stratifications obtained by the inversion code for the flare (red) and post-WLF (blue) phase; the dashed lines show the error margin determined by the inversion code SIR.} 
 \label{profile_comparison}
\end{figure*}

As shown in Fig.~\ref{profile_comparison}, the inversion code explains  the flare intensity difference by enhanced temperatures at all optical depths, i.e. predicts also the heating of the solar photosphere around $\log \tau =0$, where the photospheric continuum forms. For the pixel shown in Fig.~\ref{profile_comparison}, the continuum enhancement of $0.3 I^c_{QS}$ (continuum intensity of the surrounding quiet Sun) is achieved by an increase in temperature at $\log \tau =0$ by 840~K compared to the post-WLF phase. 

Such an increase in temperature at the deepest photospheric layers is unlikely for several reasons. First, the minimum of temperature stratification in any of the investigated pixels is not below $\log \tau =-0.5$. Second, the observed emission profiles are never in pure emission. There is always a slight decrease in intensity in the far wings of the Fe I 630.15~nm and 630.25~nm lines that indicates the temperature decrease above $\log \tau = 0$. Furthermore, if the temperature increase below the solar surface is real, the heating will have to be of a specific type, e.g. increase with depth because the atmospheric density significantly rises at those layers. 

The assumption that the flare atmosphere is probably not altered significantly around $\log \tau = 0$ is further supported by the semiempirical flare models F1, F2, F3, F1$^{*}$  \cite{1980ApJ...242..336M,1989SoPh..124..303M} and the FLA and FLB models for WLF  \cite{1990ApJ...360..715M}. These coincide with the quiet-Sun  model C of  \cite{1981ApJS...45..635V}  at heights $z < 0$~km, i.e. at $\log\tau>0$ \cite[Fig.~1]{1989SoPh..124..303M}. Also, models of continuum emission in a sunspot atmosphere heated by non-thermal electron beams \cite{2010ApJ...711..185C} indicate that the temperature below $z=0$~km increases not more than by 100~K for one of the strongest beam heating they used. Little or no photospheric heating was observed also in other models \cite{2017A&A...605A.125S,2016ApJ...827..101K}.

\begin{figure}
    \centering
    \includegraphics[width=\textwidth]{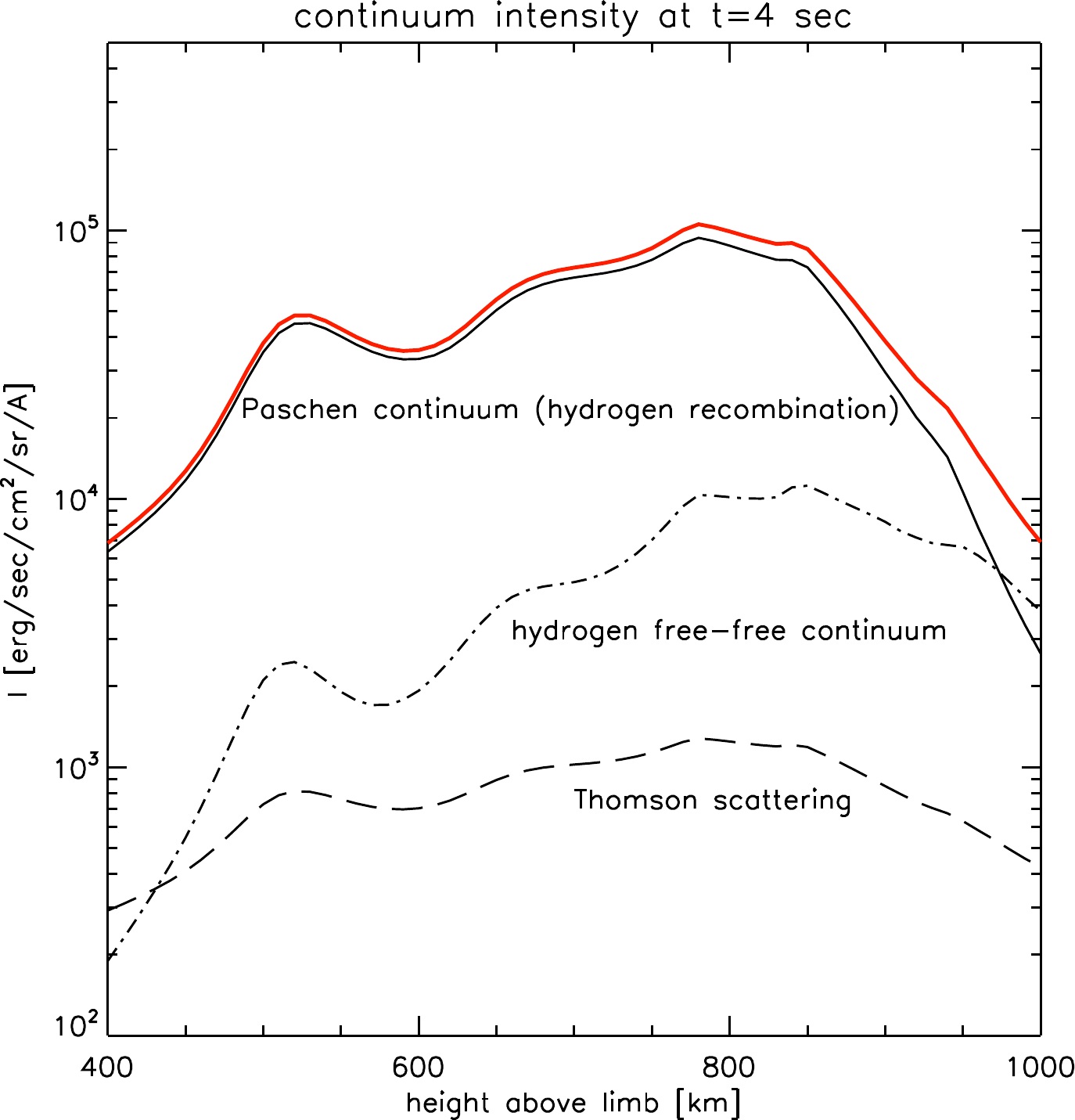}
    \caption{Vertical variations of the line-of-sight contributions to the continuum intensity from FLARIX radiation-hydrodynamical simulation at time $t=4$~s. The total continuum emission (red line) is dominated
    by the Paschen continuum (full black line) with minor contributions from the hydrogen free-free (dotted-dashed line) and Thomson scattering (dashed line) components. See also  \cite{Heinzel:2017}. Courtesy of P. Heinzel.}
    \label{fig:offlimb}
\end{figure}
Moreover, the observations of off-limb visible continuum sources during a M7.7 and a M1.7 flare were recently analysed \cite{Heinzel:2017}. 
Using analytical formulae and detailed radiation-hydrodynamical simulations, the authors 
conclude that the dominant source of the off-limb visible continuum radiation is  the Paschen recombination continuum and there are also smaller contributions by  Thomson scattering and by the hydrogen free-free emission, see Fig.~\ref{fig:offlimb}.
Futhermore, both HMI observations and numerical simulations show, see Fig.~1 in  \cite{Heinzel:2017} and Fig.~\ref{fig:offlimb}, respectively,  that the continuum radiation from these sources is located in the solar chromosphere, i.e. well above the line-forming region of the Fe I 630.15~nm and 630.25~nm lines. In the studied cases of M-class flares, the off-limb continuum intensity was around $0.1 I^c_{QS}$. 

Naturally, the inversion code SIR cannot account for such potential sources of  chromospheric continuum emission as it is applied to a photospheric line. Instead, SIR compensates for the continuum rise by increasing the temperature around the photospheric continuum-formation layer. To investigate whether such a contribution of continuum intensity from the solar chromosphere is realistic, a set of inversions was performed on the Stokes profiles observed in the region encircled by the blue contour in Fig.~\ref{hinode_int}. In order to mimic such a chromospheric optically-thin contribution, the Stokes $I$ intensity was artificially decreased by a flat continuum, where for each pixel the observed WLF continuum was decreased  to the continuum intensity observed in the post-WLF phase (the step of the decrease was $0.01 I^c_{QS}$). This fine step of decreasing the $I^c_{QS}$ was used to  find the best match of temperature stratifications for the flare and post-WLF phases at each pixel, and  does not necessarily mean a matching continuum intensity. Also, it is assumed that the continuum contribution from the chromosphere is unpolarised, i.e. the Paschen continuum; therefore, the Stokes $Q$, $U$, and $V$ profiles remained the same.

\begin{figure*}[!t]
 \sidecaption
 \includegraphics[width=0.9\linewidth]{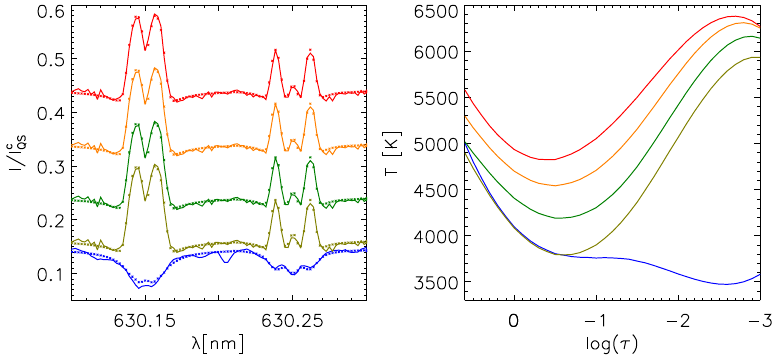}
 \caption{Comparison of the temperature stratifications resulting from the inversion of Stokes profiles, where the Stokes $I$ profile was reduced by a flat continuum of $0 I^c_{QS}$ (red), $0.1 I^c_{QS}$ (orange), $0.2 I^c_{QS}$ (green), and $0.28 I^c_{QS}$ (olive). The blue Stokes $I$ profile corresponds to the post-WLF phase at the same location and the temperature stratification indicated by the blue line corresponds to this profile. Left: Lines correspond to the observed profiles, and the $*$~symbols  to the best fits obtained by the inversion code SIR.} 
 \label{profile_reduction}
\end{figure*}

In Fig.~\ref{profile_reduction} the results of these inversions are displayed for the pixel shown in Fig.~\ref{profile_comparison}, where the red and blue lines are identical in these plots. It is clear that the inversion code SIR can also reliably fit  Stokes profiles when the Stokes $I$ profile is artificially decreased by a flat continuum. Examples of such decreased Stokes $I$ profiles are shown in the left plot in Fig.~\ref{profile_reduction}; the  corresponding temperature stratifications are shown in the right plot. The other physical parameters of the model atmosphere (such as magnetic field structure) are not affected significantly by the modification of the Stokes $I$ profile. 

Next, the best match is sought between model temperatures obtained from the inversions of the reduced Stokes $I$ profiles and the post-WLF temperature stratification in the same pixel in the interval of $\log\tau=[0.3,-0.3]$. For the case of the pixel shown in Fig.~\ref{profile_reduction}, the best match is displayed by olive lines, and the temperature stratifications  match within their uncertainties  in the range of optical depths from $\log \tau =0.7$ to $-0.7$. This is the range of optical depths where most of the photospheric continuum  forms.
 
As implied before, the reduction of the Stokes $I$ profile by a flat continuum is optimised to achieve the match of the temperature value at $\log \tau =0$ for the flare and post-WLF phases. This does not necessarily mean that the continuum intensity would be the same for the reduced Stokes $I$ profile and the profile observed during the post-WLF phase. In the case of the pixel shown in Fig.~\ref{profile_reduction}, a discrepancy of $0.02 I^c_{QS}$ is found between the continuum intensity of the olive-coloured emission profile and the blue post-WLF profile. This is caused by a small contribution to the continuum intensity from the upper layer of the solar photosphere that is heated during the flare phase. 

\begin{figure*}[!t]
 \centering \includegraphics[width=0.85\linewidth]{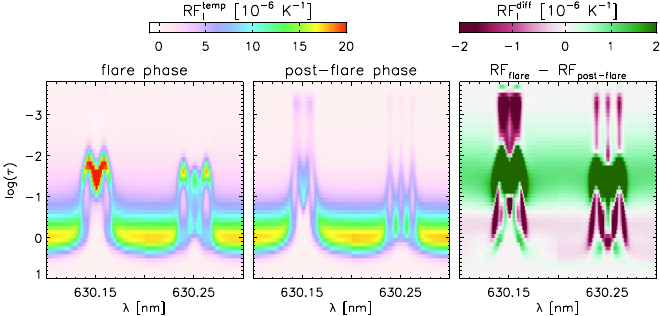}
 \caption{Response functions of the Stokes $I$ profile to the temperature for the model atmosphere fitting the emission profile shown in Fig.~\ref{profile_reduction} (left panel, in olive green) and for the model atmosphere fitting the post-WLF profile  in Fig.~\ref{profile_reduction} (middle panel, in blue). Their difference is shown in the right panel.} 
 \label{rf}
\end{figure*}

This is clearly seen in Fig.~\ref{rf}, where the response functions of the two compared Stokes $I$ profiles to the changes in their respective temperature stratifications are shown. The difference between these response functions (shown in the right panel of Fig.~\ref{rf}) indicates that  in the case of the emission profile, the photospheric continuum also forms  in higher photospheric layers, mostly between $\log \tau =-1$ and $-2$. 

The spectral-line inversions of the photospheric line thus allowed us to infer some information about the higher levels of the atmosphere. First of all, it is clear that the flare modifies the atmosphere significantly. Second, from the presented modelling of the \emph{photosphere} it turned out that there very likely is a layer in the \emph{chromosphere} or above, where a spectral continuum contribution forms during the flare event. 

\section{Summary}

In the two examples we demonstrated that the visible-range electromagnetic radiation originating in the thin layer of the solar photosphere actually may contain also information about the adjacent layers of both the solar interior and the solar atmosphere. This information must however be sought out by using sophisticated methods. In the two examples we showed the main methodology was that of the mathematical inversion. 

The framework of the two is very similar. Both methods rely on existence of functions having a meaning of Fr\'echet derivatives of the functional with respect to the free parameters of the model. In the case of local-helioseismology inversions these functions are termed sensitivity kernels, in the case of spectral-line inversion they are called response functions. Their meaning is very much the same. They represent a linear response of the perturbation in the free parameters of the model (flow and speed of sound in case of local helioseismology, atmospheric stratification in the case of spectral-line inversions) in the observables (travel times of the waves in the case of local helioseismology, the Stokes vector in the case of the spectral-line inversion). 

Both problems end up with construction of the cost function which is minimised with respect to the free parameters of the model. The exact mathematical method how this is solved is different. Whereas the local helioseismology relies on performing the matrix inversion and thus the resulting model modifications are valid strictly only in a linear regime, in the case of spectral-line inversions the process is iterative, where the linear relation is considered in each iteration separately. Thus, in the end many linear iterative step may lead to a final change of the starting model which is far beyond the linear approximation from the initial model. 

It is very difficult in both cases to ensure that the obtained solution is the correct one. In the case of inversions for local helioseismology, the solution may be influenced by a choice of the trade-off parameters. Thanks to the fact that the output of the inversion is not only the inverted estimate of the free parameter, but also the localisation (averaging) kernel and the estimate of the noise level, all these three quantities should allow for a proper interpretation of the results regardless of the choice of the trade-off parameters. A proper selection of their values is far beyond the scope of this review and belongs to the ``art of helioseismology''. In the case of the spectral-line inversion, the goodness of the fit is described purely by the $\chi^2$ value, from which the uncertainties of the estimates of the free parameters may be evaluated. It is not possible to ensure that the iterative process reaches the global minimum. It is thus important to select a proper starting atmospheric model which helps the convergence process. A selection of the best starting model is very peculiar and might also be termed with a word ``art'' and is influence also by the experience of the user knowing the code and its behaviour. 


\begin{acknowledgement}

The authors were supported by the Czech Science Foundation under grants 18-06319S (M.\v{S}., J.J. and D.K.) and 19-09489S (J.K.). J.J., J.K., and M.\v{S}. acknowledge the support from the project RVO:67985815. D.K. is supported by the Grant Agency of Charles University under grant No. 532217. We thank Juraj L\"orin\v{c}\'ik and Petr Heinzel for useful comments on an early draft of the manuscript and the referee for constructive comments. 
\end{acknowledgement}

\bibliographystyle{spphys}
\bibliography{references.bib}
\end{document}